# A community-based transcriptomics classification and nomenclature of neocortical cell types

Rafael Yuste[1], Michael Hawrylycz[2], Nadia Aalling[3], Detlev Arendt[4], Ruben Armananzas[5], Giorgio Ascoli[5], Concha Bielza[6], Vahid Bokharaie[7], Tobias Bergmann[3], Irina Bystron[8], Marco Capogna[9], Yoonjeung Chang[10], Ann Clemens[11], Christiaan de Kock[12], Javier DeFelipe[13], Sandra Dos Santos[14], Keagan Dunville[15], Dirk Feldmeyer[16], Richárd Fiáth[17], Gordon Fishell[10], Angelica Foggetti[18], Xuefan Gao[4], Parviz Ghaderi[19], Onur Güntürkün[20], Vanessa Jane Hall[3], Moritz Helmstaedter[21], Suzana Herculano-Houzel[14], Markus Hilscher[22,23], Hajime Hirase[24], Jens Hjerling-Leffler[23], Rebecca Hodge[2], Z. Josh Huang[25], Rafiq Huda[26], Yuan Juan[9], Konstantin Khodosevich[3], Ole Kiehn[3], Henner Koch[27], Eric Kuebler[28], Malte Kühnemund[29], Pedro Larrañaga[6], Boudewijn Lelieveldt[30], Emma Louise Louth[9], Jan Lui[31], Huibert Mansvelder[12], Oscar Marin[32], Julio Martínez-Trujillo[28], Homeira Moradi[33], Natalia Goriounova[12], Alok Mohapatra[34], Maiken Nedergaard[3], Pavel Němec[35], Netanel Ofer[36], Ulrich Pfisterer[3], Samuel Pontes[1], William Redmond[37], Jean Rossier[38], Joshua Sanes[10], Richard Scheuermann[39], Esther Serrano Saiz[1], Peter Somogyi[8], Gábor Tamás[40], Andreas Tolias[41], Maria Tosches[21], Miguel Turrero Garcia[10], Argel Aguilar-Valles[42], Hermany Munguba[23], Christian Wozny[43], Thomas Wuttke[27], Liu Yong[9], Hongkui Zeng[2], Ed S. Lein[2]

1 Columbia University
2 Allen Institute for Brain Science
3 University of Copenhagen
4 European Molecular Biology Laboratory
5 George Mason University
6 Universidad Politécnica de Madrid
7 Max Planck Institute for Biological Cybernetics
8 University of Oxford
9 Aarhus University
10 Harvard University
11 Humboldt-Universität zu Berlin
12 VU University Amsterdam
13 Instituto Cajal (CSIC)
14 Vanderbilt University
15 Scuola Normale Superiore, sede di Pisa
16 JARA-Brain Institute of Neuroscience and Medicine
17 Hungarian Academy of Sciences
18 Christian-Albrechts-University Kiel
19 École polytechnique fédérale de Lausanne (EPFL)
20 Ruhr University Bochum
21 Max Planck Institute for Brain Research
22 Stockholm University
23 Karolinska Institutet
24 RIKEN Center for Brain Science
25 Cold Spring Harbor Laboratory
26 Massachusetts Institute of Technology
27 University of Tübingen

28 University of Western Ontario
29 CARTANA
30 Leiden University Medical Center
31 Stanford University
32 King’s College London
33 Krembil Research Institute
34 University of Haifa
35 Charles University
36 Bar Ilan University
37 Macquarie University
38 Sorbonne University
39 J. Craig Venter Institute
40 University of Szeged
41 Baylor College of Medicine
42 Carleton University
43 University of Strathclyde

**To understand the function of cortical circuits it is necessary to classify their underlying cellular diversity. Traditional attempts based on comparing anatomical or physiological features of neurons and glia, while productive, have not resulted in a unified taxonomy of neural cell types. The recent development of single-cell transcriptomics has enabled, for the first time, systematic high-throughput profiling of large numbers of cortical cells and the generation of datasets that hold the promise of being complete, accurate and permanent. Statistical analyses of these data have revealed the existence of clear clusters, many of which correspond to cell types defined by traditional criteria, and which are conserved across cortical areas and species. To capitalize on these innovations and advance the field, we, the Copenhagen Convention Group, propose the community adopts a transcriptome-based taxonomy of the cell types in the adult mammalian neocortex. This core classification should be ontological, hierarchical and use a standardized nomenclature. It should be configured to flexibly incorporate new data from multiple approaches, developmental stages and a growing number of species, enabling improvement and revision of the classification. This community-based strategy could serve as a common foundation for future detailed analysis and reverse engineering of cortical circuits and serve as an example for cell type classification in other parts of the nervous system and other organs.**

## Anatomical and physiological classifications of cortical cell types

After more than a hundred years of sustained progress by generations of neuroscientists, it is clear that neocortical neurons and glial cells, like cells in any tissue, belong to many distinct types. Different cell types play discrete roles in cortical function and computation, making it important to characterize and describe them accurately and in their absolute and relative numbers. However, classifications of cortical neurons beyond the largest divisions have generally been subjective, based on a qualitative assessment of the morphology of a small number of neurons by individual investigators. Towering historical figures like Cajal, Lorente de Nó and Szentágothai, among others, proposed classifications of cortical cells based on their morphologies as visualized with histological stains[4,6,7] **(Figure 1A-C)**. These anatomical classifications described several dozen types of pyramidal neurons, short-axon cells and glial cells, which were subsequently complemented by morphological accounts of additional cortical cell types by many researchers[8-13] but without arriving at a clear consensus as to the number or even the definition of a cortical cell type. In particular, there is no established convention for assessing which morphological features of a neuron are essential to characterize a given cell type.

Over the last few decades, the introduction of new optical microscopy, morphological, ultrastructural, immunohistochemical, and electrophysiological methods, and the widespread use of molecular markers (**Figure 1D-H**), have provided increasingly finer phenotypic measurements of cortical cells and enabled new efforts to classify them more quantitatively, using supervised or unsupervised machine learning methods such as cluster analysis [16-18]. A community effort to classify neocortical inhibitory cells was attempted at the 2005 Petilla

Convention, held in Cajal's hometown in Navarre, Spain, and led to the adoption of a common standardized terminology to describe the anatomical, physiological and molecular features of neocortical interneurons[5]. While useful, this attempt fell short of providing a classification and working framework that investigators could incorporate into their research. But, at the same time, an outcome of the Petilla Convention was the realization that there was not yet a single classification method that both captured the inherently multi-modal nature of cell phenotypes and could serve as a standard. While most researchers accepted the existence of cell types that could be captured and defined independently by different methods, there was no agreement as to which would form an optimal basis for classification. In principle, many criteria can be used, including 1) an anatomical or connectivity-based classification[21,22], 2) a parametrization of the intrinsic electrophysiological properties[19], 3) a combination of structural and physiological criteria[20,23], 4) molecular markers detected with antibodies or single cell PCR[16,24], 5) identification of developmental[25] or epigenetic attractor states[26] or 6) using evolutionary approaches that identify homologous cell types across species[27]. Ideally, all these classifications would converge and agree with each other, or at least substantially overlap. Indeed, there is indeed substantial concordance among categories based on anatomical, molecular and physiological criteria[23,28-31], but it has not been easy to combine these approaches and datasets into a unified taxonomy. There are significant differences between experts in assigning neurons to particular classes in the literature[21] and experts often disagree on what constitutes ground truth. This uncertainty is exacerbated by technical problems: conventional approaches have been laborious, low-throughput, frequently non-quantitative, and generally plagued by an inability to sample cells in non-biased ways. Thus, setting aside debates about the importance of various criteria and the nature or even existence of discrete cell types, it is not surprising that the cell type problem remains challenging. However, a new approach now available, single-cell transcriptomics, can help break the impasse.

**Transcriptomics as the core framework for classifying cell types**

Recent advances in high-throughput single-cell transcriptomics (scRNAseq) have dramatically changed the paradigm of cellular classification, offering a powerful new quantitative genetic framework for classifying cell types[32-34]. This new approach measures the expression of thousands of genes (transcriptomes) in large numbers of single cells and operates at relatively high speed and low cost. Related methods in epigenomics can identify sites of methylation and putative gene transcriptional regulation, essential to cell function and state. These new approaches descend from the methodological, conceptual and economic revolution created by the *Human Genome Project* [35,36]. With diverse genomes in hand, it became feasible to generate entire transcriptomes from tissues, and these methods were then miniaturized for amplifying the RNA present in single cells. Initially, practical considerations limited this application to only a few hundred cells per experiment but effective new methods quickly emerged for profiling thousands of cells or nuclei at a time[37] [38-41]. With simultaneous advances in computational methods needed to analyze an overwhelming amount of sequence-based data [42,43], it is now possible to classify and characterize the complete diversity of neural cells in an unbiased way in any tissue or species, including the neocortex [1,33].

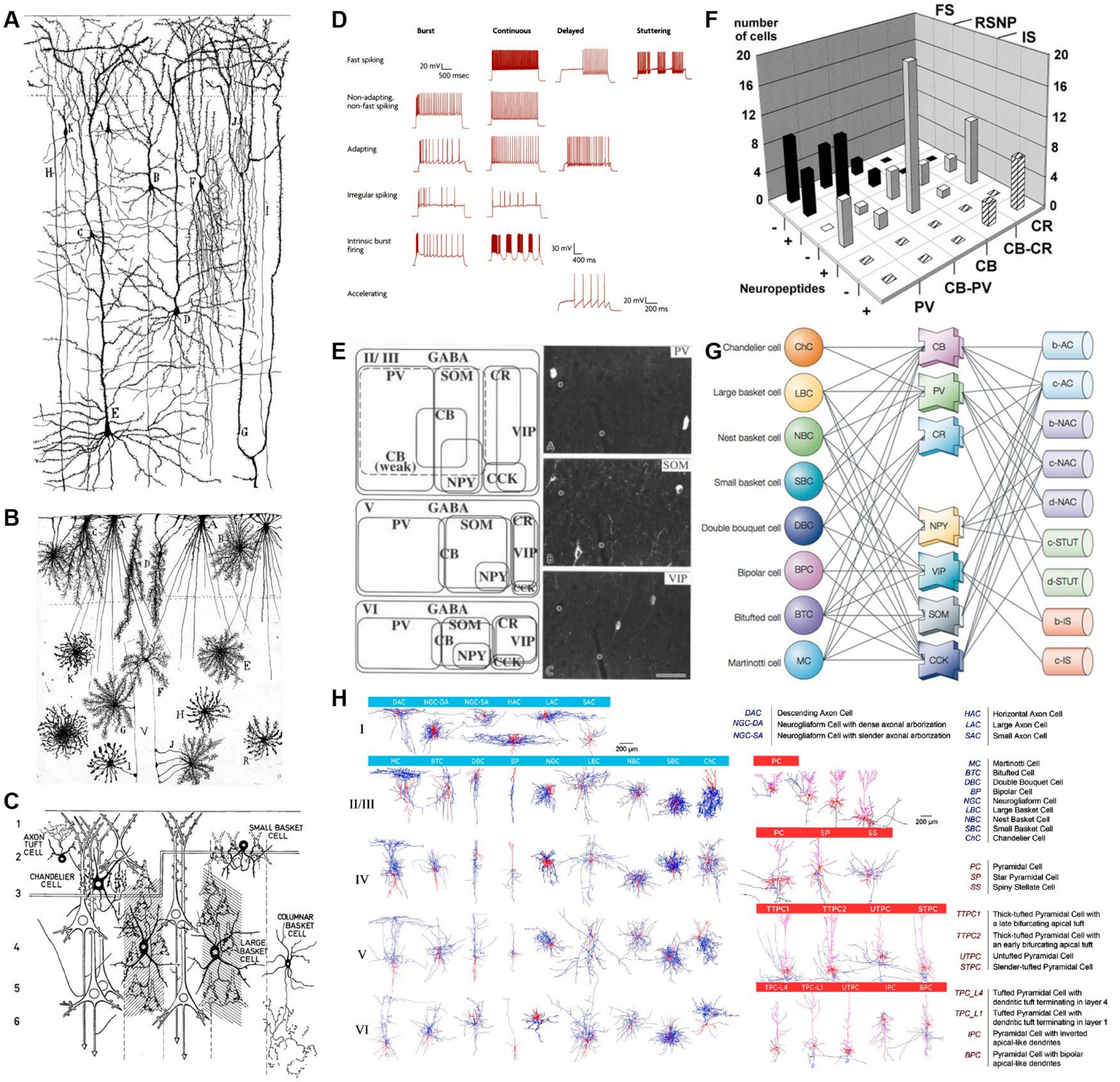

***Figure 1: Historical milestones in cortical cell type classification.*** *Morphological characterization and classification of neurons (**A**) and glial cells (**B**) by Ramón y Cajal (1899).[3]* ***C***. Diagram showing the connections of different types of interneurons with pyramidal cells; from *Szentágothai* (1975)[4]. ***D****. Cortical cell type classifcation based on intrinsic firing properties (from Petilla convention)[5].* ***E****. Definition of GABAergic interneuron classes based on non-overlapping and combinatorial marker gene expression; from Kawaguchi and Kubota (1997)*[14]. **F**. *Correlation of firing properties with class markers.* ***G****. Complex relationships between cellular morphology, marker gene expression and intrinsic firing properties based on multimodal analysis (from Markram et al. (2004)*[19]*).* ***H****. Comprehensive morphological and physiological classifications of cortical cell types (from Markram et al. (2015)*[20]).

Conceptually, as much as the genome is the internal genetic description for each species, the transcriptome, as the complete set of genes being expressed, provides another internal code that describes each cell within an organism in a spatiotemporal context. Practically, the scale of scRNAseq promises near-saturating analysis of complex cellular brain regions like the neocortex or any other organ, providing for the first time a *comprehensive* and *quantitative* description of cellular diversity and the prospect of simplifying tissue cell composition to a finite number of cell types and states defined by the clustering of these datasets. Importantly however, these transcriptionally-defined clusters represent a probabilistic description of cell types in a high-dimensional landscape of gene expression across all cells in a tissue, rather than a definition based on a small set of necessary and sufficient cellular markers or other features.

The scale, precision and high information content of these current methods now far outpace other classical methods of cellular phenotyping in neuroscience and have the potential to approach the criteria of Complete, Accurate and Permanent (CAP) often cited by the late Sydney Brenner as the gold standard in science [44,45]. Indeed, major consortium efforts now aim to generate a complete description of cell types based on molecular criteria across the cortex (*Allen Institute for Brain Science*[1,33]), the entire brain (the *NIH BRAIN Initiative Cell Census Network* [46] and even the entire body (the *Human Cell Atlas*[47]). As the Human Genome Project offers a means for comparative analysis of orthologous genes across species, these efforts promise to define all or most cell types and states in humans and model organisms, with the possibility of extending them to a variety of species to understand the evolution of cell type diversity. These enormous investments have the potential for a transformative effect on the neuroscience community, which will be accelerated by a formalization of a molecular classification and adoption by the community.

Transcriptomic classifications offer a number of advantages when used as a framework for bounding the problem of cellular diversity[47,48]. For example:

1) High-throughput transcriptomics is uniquely effective at allowing a systematic, comprehensive analysis of cellular diversity in complex tissues. Its quantitative and high-throughput nature enables the adoption of rigorous definitions and criteria using datasets from tens of thousands to millions of cells.

2) The genes expressed by a cell during its developmental trajectory and maturity ultimately underlie its structure and function, and so the transcriptome offers predictive power based on interpreting gene function. From this perspective, other cellular phenotypes, including morphology, are in part encoded by genes, rather than completely independent defining criteria[49]. Of course the transcriptome of the mature neuron, measured at a single point in time, does not fully predict cellular properties for many reasons: it fails to capture the cell's developmental history, with intrinsic or extrinsic influences determining its phenotype (such as interaction within its synaptic microcircuitry or neuromodulatory effects), nor does it reveal post-transcriptional or post-translational modifications, regulation, trafficking or physiological brain-state dependent relatively short term modulation of subsets of genes. Nevertheless, in general, transcriptome-based classifications are so far largely concordant with a large body of literature regarding cellular anatomy, physiology, epigenetic markers, function and developmental origin, while offering an open-ended means for generating hypotheses about gene expression underlying other cellular phenotypes.

3) A molecular definition of cell types allows the identification of robust cell type markers and the creation of genetic tools to target, label and manipulate specific cell types [50,51]. Even if such tools do not resolve the lumping or splitting of discrete subtypes, they provide the means to standardize the datasets obtained by different researchers.

4) Transcriptomic data also can provide information about human diseases, allowing a potential linkage between genes associated with disease and their cellular locus of action. Admittedly, pathological disturbances could be deduced, independent of transcriptomics, from pathophysiological studies combined with time and place-dependent modifications. The cell type transcriptomics-based data might lead to identification of many mechanistically unresolved diseases, as changes in the expression levels of key genes from involved cell types.

5) And finally, the transcriptome is unique among cellular phenotypes in that it allows quantitative alignment of cell types between highly disparate datasets based on conserved molecular signatures across evolutionary or developmental time. Indeed, advances in single nucleus sequencing now allow transcriptomic analysis in any species or developmental stage, enabling the potential for alignment of cell types across species (based on conserved expression of homologous genes) and developmental stages (based on gradual developmental trajectories) [2,52]. Systematic cross-species comparisons of similarly acquired and analyzed single-cell transcriptomes will also make possible the objective examination of the degree to which cell type diversity in the cerebral cortex has increased or been constrained in evolution.

Proposing a molecularly based classification scheme for use by a field traditionally centered on cellular anatomy, physiology and synaptic connectivity may be challenging unless such a classification correlates strongly with those features. Recent work in the retina is promising in this regard, where a large body of work has established a highly diverse set of anatomically, physiologically and functionally discrete cell types[53]. For example, for mouse bipolar cells, a class comprising 15 types of excitatory interneurons, there is essentially a perfect correspondence between types defined by scRNAseq, high-throughput optical imaging of electrical activity, and serial section electron microscopy [32,54-57]. Application of single- cell transcriptomics to the retina identifies clusters that strongly correlate with this prior cell type knowledge [32,53,58]. Whereas an analogous concordance may not be as straightforward in the neocortex, to date there is little evidence that it cannot be achieved.

Importantly, transcriptomics results can be complex and, even within a single putative cell type, there could be variation in gene expression due to cell state, differentiation, and other dynamic processes. Some studies have suggested that cell types are less defined discrete entities but rather are components of a complex landscape of possible states [59-61]. Further, there are aspects of genome regulation such as transcription-factor binding activities, that are not yet measurable in single cells [62]. Differentially regulated genes that establish these cell states or drive transitions will need to be identified and analyzed. In this respect, the extensions of scRNAseq to epigenomics methods measuring open chromatin and methylation state are now becoming possible and will help to understand the impact of gene regulation on cell state [63]. Progress in the simultaneous measurement of transcription and regulatory state will continue, and having a solid transcriptome-based taxonomy will form a strong foundation for further elucidation of cell type characteristics in different mammalian cortices.

Experimental tools are increasingly available to aid in transcriptomic classification and phenotypic characterization in model animals, such as specific *Cre* lines and viruses, Patch-seq and spatial transcriptomic methods such as MERFISH. These efforts will be best approached as a community effort, linking evidence derived by them to a molecular framework. Of course it is possible that there will be significant mismatches between phenotypes, as revealed by different methods, that will be problematic for the usefulness of a molecular classification; for example, long-range connectivity patterns may have been set up early in development and may not be correlated with adult gene expression. This information will need to be incorporated into a cell type classification. Potential mismatches, however, do not negate the value of a core transcriptomic classification. Genes differentiating types are likely to have cellular functions, genes are the linkage to genetics of human disease, and genes are the only path to genetic tools to manipulate cell types.

**Biological insights from cortical transcriptomics**

A transcriptomics classification is not only practical but could also enable important biological discoveries. Indeed, the application of scRNAseq to mouse and human cortex has already identified a complex but finite set of molecularly defined cell types that generally agree with the vast prior literature on cytoarchitectural organization, developmental origins, functional properties and long-range projections[64]. Moreover, these initial transcriptomic studies of cortical tissue are proving their meaningfulness by providing key insights into the biology of the system. To start, the hierarchical (agglomerative) organization of transcriptomic cell types, based on relative similarity between clusters, makes strong biological sense. Viewed as a tree or dendrogram, the initial branches reflect major classes (neuronal vs. non-neuronal, excitatory vs. inhibitory), with finer splits reflecting more subtle variants of each class. The major splits likely reflect different developmental programs; for example, neocortical neurons are split into excitatory glutamatergic vs. inhibitory GABAergic classes reflecting their different developmental origins in embryonic pallium versus subpallial proliferative regions, while the next splits in the GABAergic branch contain neurons generated by medial and caudal subdivisions of the ganglionic eminences and the preoptic areas (**Figure 2A)**. These transcriptomic divisions are consistent with a long literature on cell fate specification of different GABAergic classes and the transcription factors involved in that process (**Figure 2B**). Transcriptomics also allows an quantitative analysis of developmental trajectories involved in this specification and maturation[2] (**Figure 2C**). Finally, the genes that differentiate different transcriptomic classes are predictive of their cellular and circuit function, as differential expression of genes associated with neuronal connectivity and synaptic communication define these classes [34](**Figure 2D**).

Determining how consistent the organization of transcriptomic cell types is across a wider range of species will contribute to ratify the validity of this classification – or to understand the biological and evolutionary origins of its limitations. Transcriptomic classification provides a direct avenue for quantitative comparative analysis across species, by aligning cell types across species based on shared gene covariation. This is particularly relevant to better understand the human brain, because this alignment of cell types allows inference of cellular phenotypes in the human brain where such information is extremely difficult or impossible to obtain. For example, a recent study of human cortex[65] demonstrated that the overall cellular organization of the human cortex is highly conserved with that of the mouse, allowing identification of homologous cell types. Similarly, a recent transcriptomic study performed in the mouse, turtle and lizard found that the same major classes of cortical GABAergic neurons (somatostatin, parvalbumin-like and serotonin receptor 3A HTR3A) exist in mammals and reptiles[15] **(Figure 2E)**. Despite the overall conservation between mouse and human, many differences are seen in homologous types, including their proportions, laminar distributions, gene expression, and morphology (for example, *HTR3A* is not expressed in human GABAergic interneurons [65]). Another recent comparative transcriptome study of prefrontal cortex of humans, chimpanzees and macaques revealed that, in spite of its apparent histological conservation, the human neocortex shows many unique gene-expression features [66]. Finally, there are also new insights for non-neuronal cells from transcriptomic studies, which have identified astrocyte diversity and divergent molecular phenotypes between mouse and human that correlate with known morphological specializations in primate astrocytes [33,67,68].

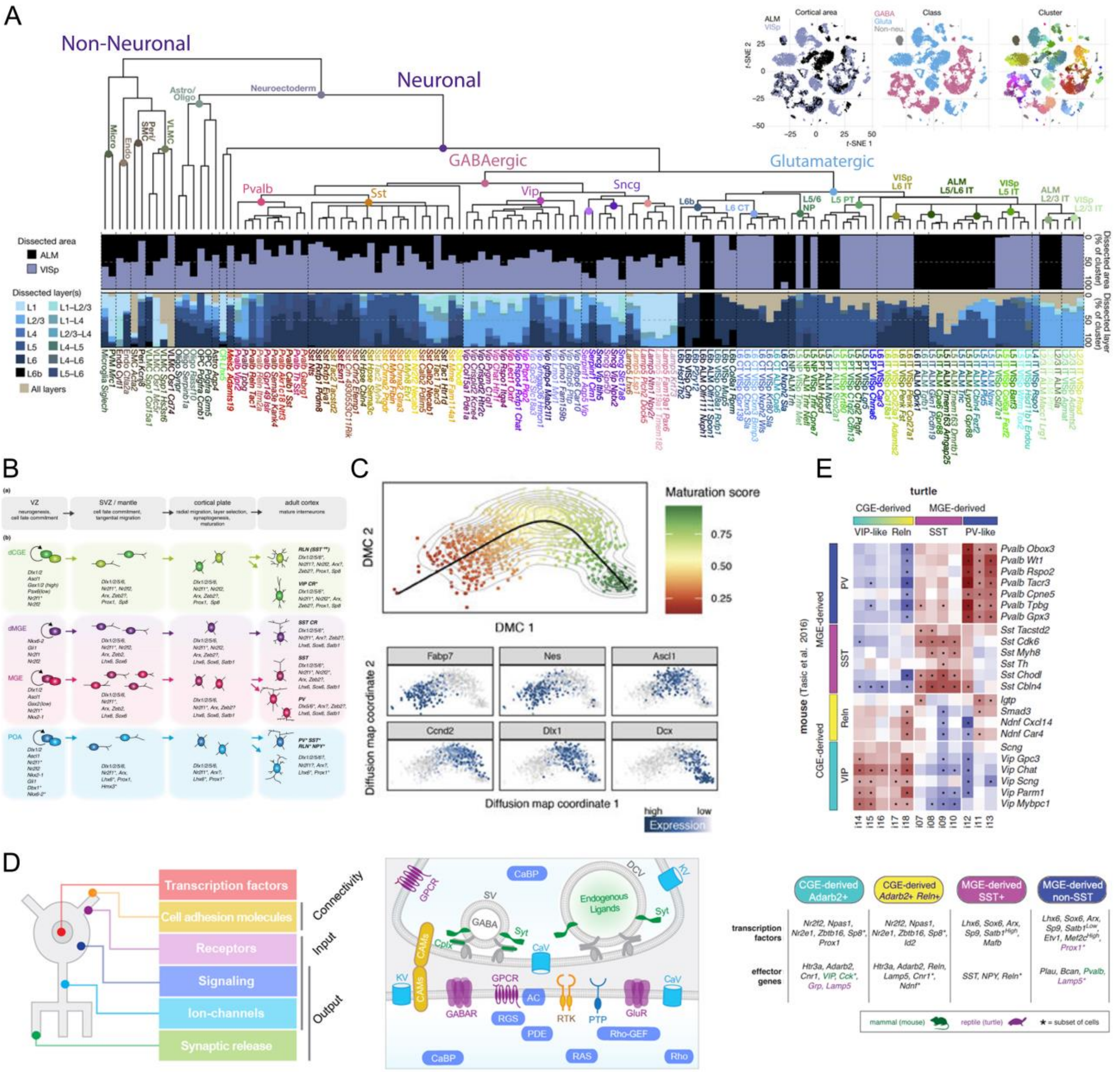

***Figure 2: Biological insights from molecular analyses of cortical cell types.*** *A. Comprehensive single-cell transcriptome analysis reveals molecular diversity of cell types, with relatively invariant interneuron and non-neuronal types across cortical areas but significant variation in excitatory neurons (modified from Tasic et al. (2018))*[1]*. B. Major interneuron classes are specified by distinct transcription factor codes (from Kessaris et al. (2014). C. Single-cell transcriptomics of GABAergic interneuron development demonstrates gradual changes in gene expression underlying developmental maturation (modified from Mayer et al. (2018))*[2] *D. Gene families shaping cardinal GABAergic neuron type include neuronal connectivity, ligand receptors, electrical signaling, intracellular signal transduction synaptic transmission, and gene transcription. These gene families assemble membrane-proximal molecular machines that customize the input–output connectivity and properties in different GABAergic neuron types. E. Single-cell transcriptomics allows cross-species comparisons and demonstrates conservation of major cell classes from reptiles to mammals, with conserved transcription factors but some species specific effectors (from Tosches et al. (2018) and Tosches and Laurent (2019))* [15]*.*

**A probabilistic definition of cortical cell types**

While there is compelling evidence for the existence of distinct cell types based on robust clusters of observable and measurable cell attributes, a precise definition of a type is more challenging as different and partially conflicting classifications have been put forward, emphasizing structural/functional characteristics [48] or cellular identities [69,70]. In an effort to arrive at a *conceptual* definition of cell types, many different criteria have been proposed. For example, a cell type may consist of groups of neurons that share a common developmental origin, common sets of gene expression patterns (such as a necessary and sufficient transcription factor code), common morphological or physiological features, or a common function in the synaptic circuit, either through input-output connectivity or the transfer function that they carry out within a same environment, while processing their inputs. While each of these views has merit and should be ultimately explained by a meaningful definition of cell type, they are often not all readily measurable nor easily combined, particularly acrosss species.

Given the complexity of cellular function, the present lack of full correspondence of multimodal data sets within and across species, and the required explanatory power that a meaningful definition of cell type should support, a plausible way forward may be to utilize an *operational* definition of a cell type, building on the existence of statistically defined clusters over a set of measurable attributes. Indeed, in most recent single-cell sequencing approaches, groups of transcriptomic profiles are clearly identified by data clustering, whereby sets of cells are subjected to a variety of iterative and hierarchical clustering methods with subsequent permutation testing for significance. As is common with basic statistical analysis, identified groups are compared to a null hypothesis of no group structure [23,71].

A critical and challenging question is how to represent a transcriptomic taxonomy. One natural approach is to adopt a hierarchical framework, a task to which cluster analysis is well suited. This approach follows the historical tradition of using cladistics to classify organisms, assuming common ancestors in their evolution and with synapomorphies (shared derived traits) among related clades. While statistical clusters do not presume any hierarchy in the structure of the data, biological systems have a temporal evolution which is one of their essential features [70]. Indeed, evolutionary history or development of a neural circuit implies earlier stages which are often less specialized and represent common ancestors of later states, making natural the structured classification of cell types as a hierarchical tree[72]. A hierarchical classification of cell types is thus particularly suited both to establishing cell type homologies across species (revealed through shared hierarchical clustering across species) and revealing species-specific cell types through objective criteria. While a hierarchical organization appears to mirror developmental principles and spatiotemporal organization in the neocortex, this may not be universally true for other brain regions, organs or across evolution to the same degree. In that case, complex inclusion/exclusion and probabilistic class relationships are not well represented hierarchically and may be more amenable to graph based or other set theoretic constructions.

This operational definition of cell types is particularly applicable to transcriptomic profiling, where the dimension of the underlying space is large, the variance comparatively high, and where competing approaches have largely identified common cell type patterns. Despite these successes, relatively little progress has been made on the rigorous definition of transcriptomic classes, and the description of intra- and inter-class variability. Moreover, there is often a discrepancy as to when to consider two cells as belonging to the same group, or two groups as being adequately different as to justify subdivision. This issue is compounded by the possibility that some transcriptomic clusters represent a state of a cell type (for example, a pathological, developmental or functional state), rather than a fixed cell type. In this case, a finer distinction among clusters may have gone too far. In fact, aside from its biological meaningfulness, one advantage of casting the cell type classification as a cladistic one is that the lumping/splitting distinction remaps itself as a distinction between different levels of the hierarchical tree, since one could split a group into subgroups at a lower level of the hierarchy to reflect data obtained in different physiological or developmental conditions. Indeed, the space of the transcriptomes for cortical cell types may be visualized as a complex high dimensional landscape with isolated peaks of expression for a given cell type but also valleys and gradients between more weakly defined classes, which could be described alternatively as types or states. Unfortunately, approaches to transcriptome cluster identification do not typically take this complexity into account when forming cell type classes[33,73,74]. While presently there is much work being done in visualization of data via clustering, a robust statistical framework that enables a quantitative definition of cell type (or tendency to be a type) is still missing.

The development of a rigorous probabilistic or statistical framework for cell type definitions would greatly assist with the interpretation of large data sources and in the identification of those attributes most relevant (and non-redundant) to cell type classification and function [75,76]. Ideally, this framework should provide a quantitative definition of a cell type that is *independent* of the methods used by clustering analysts and would include a description of quantiative metrics such as resolution, complexity, variability, uniqueness, and association of variables with other attributes. Within this statistical definition of cell type there are two alternative approaches to find and test the validity of clusters. One is "hard" clustering, with clearly defined borders between clusters and with each cell strictly assigned to a particular type. Alternatively, in "soft" (or "fuzzy") clustering, any given cell has a particular probability of belonging to a particular cluster. Despite the probabilistic nature, inter- and intra-cluster distance may still be defined for outcome validation. Also, projections to low dimensionality spaces to ease clustering visualization are possible. Although most cluster analyses performed today use hard definitions, we forsee that the biological diversity of cells in the nervous system, and the discontinuous variation found in many different types of data, including transcriptomics, as described by our landscape metaphor, are more flexibly captured with a probabilistic criterion. There is much opportunity for progress in deriving statistical and probabilistic models of cell type, models for describing data modality covariation, effective utilization of linkage correspondence methods between data modalities, and informatics frameworks for elucidating cell type structure and taxonomy. Ultimately, the consensus description of cell types may form a continuum taken to the cladistic limits, beginning with hard and ending with soft distinctions among cells types, with an ambiguous transition between these extremes.

**A unified taxonomy and nomenclature of cortical cell types**

Using such operational definitions of cell types, a data-driven transcriptomic classification of cortical cell types should allow the creation of a formal unified cell type classification, ontology and nomenclature system, whose principles are generalizable to any biological system. Following the genetic paradigm proposed, there are many lessons to learn from genomics. For example, the classification could be iteratively updated, refined with subsequent accumulation of data [77] such as builds of the genome or transcriptome, which changed dramatically in the early years and have become increasingly more stable. To accomplish this, a coherent principled nomenclature system is essential. Like in current gene nomenclature, many aliases exist linking cell types to commonly used terminology relating to cellular anatomy or other phenotypes. This nomenclature should be portable across species, much as current gene symbols refer to orthologous genes. For the classification to be useful like the genome has been, computational tools need to be developed to allow researchers to quantitatively map their data to this reference classification, conceptually similar to BLAST alignment tools [78] to map sequence data. In addition, this classification should aim to be an ontological taxonomy (i.e. describing the actual biological reality of the data, rather than just simply reflecting the statistical structure of the data itself), and follow hierarchical cladistics for the reasons described above. It should also aim to be a consensus classification that incorporates the richness of data accumulated by different groups, be presented in a curated output that is public, easily accessible, and with revisions managed by a curation committee of experts.

Creation of such an ontology is a serious project in data organization that can build on prior efforts in cell ontologies[79,80] as well as best practices established by the ontology development community (see Open Biomedical Ontology Foundry, http://www.obofoundry.org). One challenge is the existence of parallel hierarchies. For example, the spatial location of a cell within cortical layers or areas is a key feature typically represented in a hierarchically organized structural ontology, but this organization can be orthogonal to transcriptomic identity. While any given cortical region contains some number of transcriptomic types, it seems likely that many of these types will vary in a somewhat continuous fashion across cortical areas and possibly also across species. The ontological system therefore needs to be able to accommodate gradients. Likewise, the classification system should also have a temporal component to capture the developmental progression from progenitor cell division to a terminally differentiated state. In addition to species-specific classification, a key goal should be the creation of a comparative "consensus" classification representing the alignment of cell types across species.

Hand in hand with this taxonomy we propose the adoption of a formal standardized nomenclature, which we view as an essential step to organize the knowledge[22]. An old Basque proverb says, "a name is necessary for something to be", and a similar Chinese saying, "the beginning of wisdom is to call things by their right names", Drawing on the long anatomical tradition of the field, and taking advantage of the fact that humans are visual animals (which makes images easier to remember, as opposed to a list of marker genes), one possibility that is atttractive despite potential drawbacks [48] is to incorporate in this nomenclature older descriptive anatomical terms, when possible (such as chandelier, double-bouquet, basket, Martinotti, pyramidal, for example), to seek consistency with the vast literature on neocortical cell types. To name higher level branches of hierarchical trees, one could combine names, like in species taxonomy, with a name that describes the genus and a second one that describes the species. Whether this tradition should be followed for new undescribed cell types is debatable, and adopting ultimately a non-morphologically based nomenclature could make it more flexible, more easily applicable across species, and also compatible with other tissues outside the cortex or the brain. Given the explosion of scRNA-seq-based atlasing efforts now under way, developing a nomenclature convention that works for cortex, brain and even the whole body is an essential problem to be solved.

This taxonomy will only be useful and successful if adopted by the community. In addition to the nomenclature, a series of research tools should be developed, ideally by a community consortium, to facilitate similar experimental access to these cell types by the broader range of investigators. We envision molecular and genetic tools, such as standard sets of antibodies and RNA probes to identify key molecular markers for each cell type, as well as cell or mouse lines that are used as resources for the entire community. Statistical tools to enable direct comparisons among datasets, and mapping of new datasets to reference datasets, are essential. Finally, as described below, an open informatics backbone needs to be developed as an essential part of the taxonomy.

**A knowledge environment platform for community data aggregation**

While one can view the transcriptomic classification as an outcome of the genomics revolution, one could also bring to bear the capability of another recent technological revolution, the internet. If we start from the premise that it is unlikely that a biologically comprehensive and widely accepted cell type classification will be completed for several years, then a natural question is whether we have the necessary tools in place to support this effort. In addition to functional annotation, additional refinement of transcriptomic and computational methods, the invention of new methods for measuring structure, function and connectivity, and acquisition of massive amounts of data will all be needed. As is well known, there is tremendous inefficiency in the scientific process with respect to deriving and retaining value from data[81], and contemporary scientfic publications and presentations are at odds with a dynamic and growing body of knowledge with an intended integrated outcome. Modern internet applications and community environments are much more applicable to our goal, and there are several features of this approach worth adopting in a way that could harness the power of the community of researchers more effectively.

We propose that cell type classification and nomenclature could, in addition to its intrinsic scientific and medical value, simultaneously serve as a framework platform to accumulate data from the field, as a community effort. There are different possibilities for such a community taxonomy platform. For example, one could use an open website, where cortical cell types are listed and annotated, Wikipedia-like, by users. The site could look like a spreadsheet, with rows in the matrix representing each cell type and columns the different phenotypical features that characterize them. For example, in a given row, after a tentative name proposed for the cell type, examples of categories that could be annotated would include “alternative names”, “molecular features”, “morphological features”, “functional features”, “developmental origінins” and “comparative aspects”, among others. Although a Wikipedia-style model might be a convenient way of summarizing types and their supporting evidence, we think that a more useful scenario, and one that can more efficiently inform successive versions of the classification, is to invest in a larger effort to create a dynamic community knowledge framework.

The appropriate data structure we propose for such community platform, initially based on a transcriptomics cell type taxonomy, but then incorporating information from many sources, is a *Knowledge Graph* (**Figure 3**). A knowledge graph is a data structure where the nodes represent categories (in this case, cell types, using transcriptomics as the initial scaffold) and the links, or edges, between them represent their statistical relations (which can be expressed as conditional probabilities). This is represented in a multidimensional space, defined by the different metrics used to measure the nodes (in this case, the different attributes of data associated with each cell type)[82],[83]. Through a process of community data aggregation of both raw and metadata, this graph automatically updates itself, following conventional optimization algorithms, as new data can change the relative position and distance of nodes with respect to one another. In this fashion, this graph could serve as the backend infrastructure for the taxonomy described above, since the incorporation of new data could enable finer (or different) groupings of cells. As the data organization would be non-hierarchical, it would not by itself build the taxonomy, but enable data aggregation that would serve for the next iteration of cell type clustering. The taxonomy and nomenclature, on the other hand, would inform the graph, determining its nodes. Another version of such graph would indicate where similarities and dissimilarities occur between pairs of species. The cortical cell knowledge graph could start small, initialized by standardized transcriptomics data, and become multimodal as different types of data, such as connectomics or other CAP databases, become accessible. This standardized database could be powered by open source algorithms and managed and curated by database administrators.

The proposed cell type knowledge framework would represent a living and updatable resource that maintains an actively derived and flexible ontology of cortical cell types, benefitting from present active ontology efforts. It would be a dynamic database with query capability but it would only accept peer-reviewed published data in a standardized fashion and nomenclature, providing a common denominator for the research in the field, integrating quantitative and qualitative cell type classification and allowing for update subject to validation and review. The knowledge framework would utilize computational engines that allow new data to be compared and for users to query the current state of cell type understanding from the perspective of their new data. This neuronal classifier should be able to assign the most likely type to multi- or uni-modal datasets based on similarities to the current framework's knowledge. In addition to supporting literature reference, the dynamic framework might include online forums for scientific discussion and education. Ultimately, a cell type community knowledge framework would be a dynamic and living resource that researchers, clinicians, and educators would refer to as the benchmark resource for cell types in the brain and could also infuse a spirit of collaborative endeavour in this often competitive field.

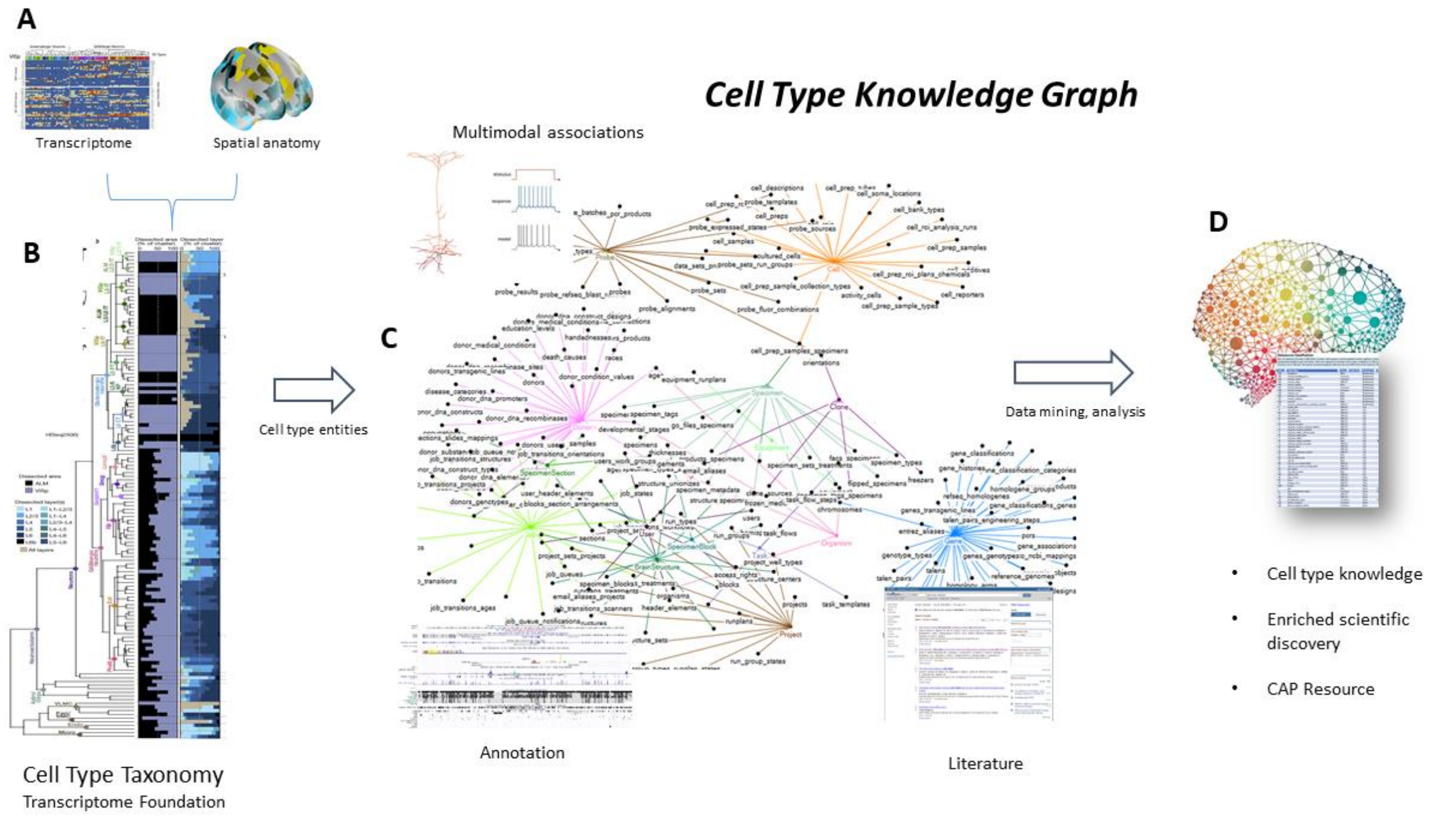

***Figure 3: Proposed community framework: taxonomy, database, and Cell Type Knowledge Graph. A**. Profiling with scRNAseq and combining transcriptomic and anatomic features yields **B**. Cell type taxonomy scaffold, forming a genetic basis for cell type characterization. **C**. Cell type measurable entities form annotations including multmodal electrophysiology, neuron morphology, epigenetic and other genome level annotations and supporting literature citations. This information is combined into a Cell Type Knowledge Graph as a graphical database supporting effective information organization. **D**. Knowledge graph architecture supports the enrichement of scientific discovery and consistent building on cell type knowledge as a Complete, Accurate, and Permanent (CAP) resource.*

## A community-based taxonomy of cortical cell types

In summary, we think that the field of neocortical studies is ready for a synthetic, principled classification of cortical cell types, based on single-cell transcriptomic data, anchored on quantitative criteria that operationally define cell clusters based on their statistical (and probabilistic) grouping, and expressed as a hierarchical tree. Although initially molecularly driven, this taxonomy should be revised and modified as other CAP datasets become available, becoming a true multimodal classification of cortical cell types. We view this core classification as potentially valid for all mammalian species, and also likely at least partly applicable to homologous structures in other vertebrates, as a broad framework to encapsulate evolutionary conservation with species specialization. Indeed, only with such a systematic approach to comparing cell types across species will it be possible to understand how cell type diversity evolved in the cerebral cortex.

In addition, we propose that the community input to support this taxonomy and enable future revisions of it is channelled into an open platform, such as a knowledge graph, as it is becoming increasingly common in community-led data science as well as the tech industry. Aggregation of knowledge through data graphs, now a common practice in the tech industry, will accelerate the dissemination of knowledge and could avoid the "publication graveyard", where data are stored away in siloed journal articles disconnected with the rest of the field. Anchoring this taxonomy and knowledge graph, a unified new nomenclature of cortical cell types valid across species is needed to centralize efforts in the field, with a generalizable framework to integrate with other cell type efforts. We view the establishment of a common nomenclature as an essential step to provide a standardized language that enables the meaningful aggregation and sharing of data.

To accomplish this effort, we envision that the classification, nomenclature and the knowledge graph could be managed by a committee of experts representing the breadth of approaches and disciplines in the field. Such a committee will be charged with designing the type of open platform to use for the knowledge graph, the statistical classification model to sustain a basic taxonomy, and the rules by which this taxonomy can be updated and revised. This committee could also decide on a unified nomenclature of cortical cells that is succinct, useful and informative, as well as methods by which community input would be incorporated in a fair and efficient fashion. This committee could arise from volunteers and would be renewed over time, flexibly representing the extent of the field and its different methodological approaches. Alternatively, this committee could be established and supported by one or a combination of already existing organizations or consortia, such as the NIH BRAIN Initiative Cell Census Network (BICCN, www.biccn.org), Allen Institute for Brain Science (www.alleninstitute.org), Human Brain Project (HBP, www.humanbrainproject.eu), Human BioMolecular Atlas Program (HuBMAP, https://commonfund.nih.gov/hubmap) or the Human Cell Atlas (HCA, www.humancellatlas.org) which are already chartered with mapping the cell types of the nervous system or other organs in the body and may have resources to build the backend IT infrastructure needed for the knowledge graph. In fact, if successful, this community-based classification effort, joined by a common nomenclature and nourished by the knowledge graph, could be extended and generalized to other systems. In this sense, the classification of neocortical cell types, a ripe field with a long tradition and multidimensional approach to a central problem in neuroscience, could be an ideal test case to explore this novel organization of knowledge in neuroscience and, more generally, in biology.

Acknowledgements**:** This document resulted from the group discussions at the FENS/Brain Prize meeting "The Necessity of Cell Types for Brain Function" that took place in Copenhagen, Denmark on October 7-10, 2018. We thank the FENS and Brain Prize Foundation and staff for help and the Lundbeck Foundation for support. This paper is dedicated to the memory of S. Brenner.